\documentclass[prc,showpacs,showkeys,nofootinbib]{revtex4}
\usepackage[dvips]{graphicx}
\usepackage{amsmath}

\begin{document}
\title{Spectral functions in the sigma-channel near the
       critical end point}
\author{Kenji Fukushima}
\email[E-mail: ]{fuku@nt.phys.s.u-tokyo.ac.jp}
\affiliation{Department of Physics, University of Tokyo, 7-3-1 Hongo,
             Bunkyo-ku, Tokyo 113-0033, Japan}
\begin{abstract}
Spectral functions in the $\sigma$-channel are investigated near the
chiral critical end point (CEP), that is, the point where the chiral
phase transition ceases to be first-ordered in the $(\mu,T)$-plane of
the QCD phase diagram. At that point the $\sigma$ meson becomes
massless in spite of explicit breaking of the chiral symmetry. It is
expected that experimental signatures peculiar to CEP can be observed
through spectral changes in the presence of abnormally light $\sigma$
mesons. As a candidate, the invariant-mass spectrum for diphoton
emission is estimated with the chiral quark model incorporated. The
results show the characteristic shape with a peak in the low energy
region, which may serve as a signal for CEP. However, we find that
the diphoton multiplicity is highly suppressed by infrared behaviors
of the $\sigma$ meson. Experimentally, in such a low energy region
below the threshold of two pions, photons from
$\pi^0\rightarrow2\gamma$ are major sources of the background for the
signal.
\end{abstract}
\pacs{05.70.Jk, 11.30.Rd, 12.38.Mh, 25.75.-q}
\keywords{sigma meson, spectral function, chiral critical end point,
 diphoton emission}
\maketitle

\section{introduction}
\label{sec:intro}

The spontaneous breaking of the chiral symmetry has fundamental
significance in understanding the non-perturbative nature of hadron
dynamics. It has been argued that the chiral symmetry can be restored
at sufficiently high temperature and/or high baryon density by means
of effective models and lattice calculations (see Ref.~\cite{kan02}
for a state-of-the-art review of lattice results). In QCD with two
massless (massive) quark-flavors at zero baryon density, the chiral
phase transition is supposed to be a second-ordered one (crossover)
according to the universality argument~\cite{pis84}. Once an adequate
chemical potential $\mu$ for the baryon density is introduced, the
chiral phase transition can become first-ordered for small quark
masses, as suggested by several model
studies~\cite{kle92,asa89,bar89,ber99,hal98}.

If the strange-quark mass is large enough to make the chiral
restoration at zero baryon density a continuous transition, as found
in the lattice calculation with staggered fermions~\cite{kan02}, the
first-ordered line should terminate at some point in the
$(\mu,T)$-plane of the QCD phase diagram. This terminal point is
called the chiral critical end point (abbreviated to CEP in this
paper). Two minima of the effective potential degenerate right at this
point, so that the curvature around the minima vanishes. This results
in the appearance of the $\sigma$ meson with \textit{zero screening
mass}, even though the pions are still massive due to explicit
breaking of the chiral symmetry. This is the reason why much attention
has been paid to physical consequences around CEP from not only the
theoretical but also the experimental point of
view~\cite{hal98,ste98,ber00,sca01,ike02,fod02}. As a matter of fact,
future experiments of the heavy-ion-collision planned in
GSI~\cite{sen02} with $\sim30\:\mathrm{A\,GeV}$ energy and in
JHF~\cite{ima01} with $\sim25\:\mathrm{A\,GeV}$ energy will explore a
lot about the high density nature of QCD, including physics around
CEP.

It would be interesting to consider the possibility to detect such
light $\sigma$ mesons directly in heavy-ion-collision experiments. The
$\sigma$ meson in a hot medium goes through such processes as
$\sigma\rightarrow2\gamma$, $\sigma\rightarrow2\pi$,
$\sigma\rightarrow2\sigma$, $\sigma\rightarrow N\bar{N}$,
\textit{etc}. It is expected that the measurement of two photons
(diphotons) can tell us prosperous information on the transient
thermal medium because electro-magnetic probes, such as leptons and
photons, hardly receives rescattering in the relatively small system
produced in a collision. Formerly the diphoton measurement was
proposed as a candidate of the implements to see the spectral changes
near the chiral restoration: The $\sigma$ mass is so reduced around
the chiral transition temperature that the $\sigma$ meson cannot decay
into two pions and thus the spectral function in the $\sigma$-channel
must be significantly
enhanced~\cite{hat85,wel92,son97,chi98,chi98b,vol98,jid01,pat02}. In
the present paper we apply the idea to the spectral changes near CEP.
Actually strong vestiges of CEP can be anticipated since the spectral
function has not just an enhanced peak but a \textit{pole}
contribution then. We are going to estimate the diphoton spectrum
taking advantage of the universality argument around CEP. However, the
information derived from the universality is not sufficient to obtain
the diphoton emission rate. For that purpose, we must take account of
each dynamical processes in order to acquire the spectral function.

In the construction of the spectral function we would emphasize the
following viewpoint: The $\sigma$ mass and the scalar meson condensate
are not sensitive to the detail of dynamical processes. Their smooth
decreases at finite temperature can be provided by almost any thermal
fluctuation. Therefore we can parametrize the behaviors \textit{a
priori} resorting to the universality argument or knowledges inferred
from model studies, as will be performed in
Sec.~\ref{sec:parametrize}. The width and the amplitude (residue of
the propagator), on the other hand, strongly depend on dynamical
processes, or in other words, depend on what kind of loops are taken
into account. Thus we must evaluate them by considering respective
processes, as will be calculated in Sec.~\ref{sec:spect}.

Once we have the spectral functions in the $\sigma$-channel, we can
investigate the effects on the diphoton emission coming from the
spectral changes near CEP. The dominant contribution from the $\sigma$
pole brings about a characteristic shape in the diphoton spectrum,
which reflects how long the system stays around CEP. In contradiction
to a naive expectation as mentioned above, we will find in
Sec.~\ref{sec:diphoton} that the $\sigma$ pole contribution is
strongly suppressed by the small residue in infrared regions. Thus the
diphoton multiplicity is weaken due to infrared behaviors of the
$\sigma$ meson. Finally we summarize our results and discuss the
possibility to detect CEP via the diphoton measurement in
Sec.~\ref{sec:summary}.

\section{parametrization near the critical end point}
\label{sec:parametrize}

\begin{figure}
\includegraphics[width=7cm]{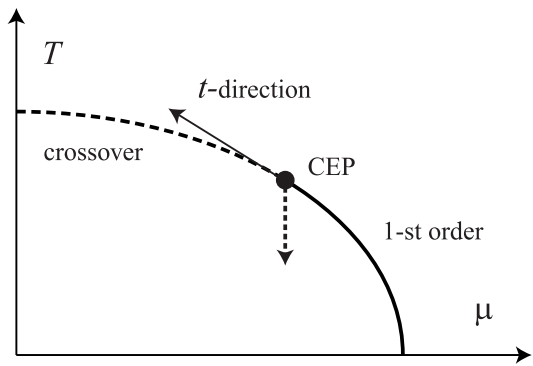}
\caption{Schematic picture of the QCD phase diagram. CEP is an
abbreviation for the critical end point, which is located at
$(\mu_{\mathrm{E}},T_{\mathrm{E}})$. The dotted line with an arrow
indicates the evolving direction considered in the present analysis.}
\label{fig:phase}
\end{figure}

In the vicinity of CEP located at $(\mu_{\mathrm{E}},T_{\mathrm{E}})$,
the critical behaviors are predominantly described by the lightest
mode, \textit{i.e.}~the $\sigma$ meson (long-ranged correlation in the
scalar-isoscalar-channel) alone. Since the $\sigma$ meson is a scalar
particle associated with the Z(2) symmetry, the universality argument
tells us that the chiral phase transition belongs to the same
universality class as that of the 3-$d$ Ising model. It is worth
noting that the chiral restoration at CEP would become a genuine phase
transition, namely, a second-ordered phase transition, even though it
is a crossover at lower baryon chemical potential. Then singularities
around that critical point are characterized by the critical exponents
as functions of the distance from the critical point. In general this
distance $l$ in the $(\mu,T)$-plane becomes an admixture of both
temperature-like (denoted by $t$) and magnetic-like (denoted by $h$)
variations in the Ising model: The Z(2) symmetry is only exact right
at CEP and the effective potential is distorted away from that
point. In the QCD phase diagram the $t$-like direction, along which
deformations of the effective potential should preserve the Z(2)
symmetry, is tangential to the first-ordered line, as depicted in
Fig.~\ref{fig:phase}.

The critical exponents for the 3-$d$ Ising model have been well
examined analytically as well as numerically~\cite{gui97}. The inverse
correlation function and the spontaneous magnetization would vanish as
$\xi^{-1}\sim t^\nu\sim t^{0.63}$ and $m\sim t^\beta\sim t^{0.33}$
along the $t$-direction and as $\xi^{-1}\sim h^{\nu/\beta\delta}\sim
h^{0.40}$ and $m\sim h^{1/\delta}\sim h^{0.21}$ along the
$h$-direction, respectively. It turns out from these numerical values
that the magnetic singularities should dominate as long as $l$
contains a non-vanishing component in the $h$-direction, or in other
words, they should dominate unless the direction specified by $l$ is
parallel to the pure $t$-direction tangential to the first-ordered
line. As a result, singularities in almost all directions can be
effectively described by the magnetic exponents, as discussed in
Ref.~\cite{ste98}. Then the mass of the $\sigma$ meson, which can be
interpreted as the inverse correlation length in the Ising model,
vanishes near CEP as $\sim l^{\nu/\beta\delta}$, while the scalar
condensate drops off as $\sim l^{1/\delta}$ as though it would be a
spontaneous magnetization. The pion mass $m_\pi$, on the other hand,
can be approximately regarded as constant up to the critical point,
which is common in several model studies~\cite{hat85,sca01,chi98}.

For simplicity we restrict our discussion throughout this paper only
to the case with the chemical potential fixed at
$\mu=\mu_{\mathrm{E}}$ (the direction indicated by the dotted line in
Fig.~\ref{fig:phase}). In this case, $l$ is proportional to
$T-T_{\mathrm{E}}$, and we can parametrize the (screening) masses
simply as follows:
\begin{equation}
 m_\pi(T)=m_\pi^\ast,\quad
 m_\sigma(T)=m_\sigma^\ast\biggl\{1-\biggl(\frac{T}{T_\mathrm{E}}
  \biggr)^2\biggr\}^{\nu/\beta\delta},
\label{eq:mass}
\end{equation}
with $\nu/\beta\delta=0.403$. Also the scalar condensate, that is, the
counterpart of the spontaneous magnetization in the Ising model can be
parametrized as
\begin{equation}
 f_\pi(T)=v_0+v(T)=af_\pi^\ast+(1-a)f_\pi^\ast\biggl\{1-\biggl(
  \frac{T}{T_\mathrm{E}}\biggr)^2\biggr\}^{1/\delta},
\label{eq:fpi}
\end{equation}
with $1/\delta=0.208$. Here $v_0$ is the condensate remaining at CEP
due to explicit breaking of the chiral symmetry. It should be noted
that the scalar condensate can be regarded as the pion decay constant
under some approximations. Accordingly we denote the condensate as
$f_\pi(T)$ in Eq.~(\ref{eq:fpi}). Indeed, this identification for the
pion decay constant is used in the choice of the parameter sets, as
stated below.

The universality argument tells nothing about the values of
$m_\pi^\ast$, $m_\sigma^\ast$, $f_\pi^\ast$ (the pion mass, the
$\sigma$ mass, and the pion decay constant at
$(\mu=\mu_{\mathrm{E}},\: T=0)$ respectively), and $v_0$. In the
present analysis these are all treated as free parameters, which
should be arranged by hand or by using some models. Here, avoiding
artifacts inherent in any model study, we simply employ the typical
parameter sets as follows, that is, we take two extreme cases, one
with no change of the $\ast$-quantities and another with a relatively
large change of the $\ast$-quantities:
\begin{align}
 &\text{(CASE-I)}\qquad m_\pi^\ast=140\;\mathrm{MeV},\quad
  m_\sigma^\ast=600\;\mathrm{MeV},\quad
  f_\pi^\ast=93\;\mathrm{MeV},\notag\\
 &\text{(CASE-II)}\qquad m_\pi^\ast=140\;\mathrm{MeV},\quad
 m_\sigma^\ast=300\;\mathrm{MeV},\quad
 f_\pi^\ast=46.5\;\mathrm{MeV}.
\end{align}
As for the choice of $a$, for the moment we will show the results of
the spectral functions and the diphoton emission rates only for a
specific choice of $a=1/2$; the condensate at CEP decreases up to the
half of its value at zero temperature. Finally we will present in
Fig.~\ref{fig:final} the results for a variety of $a$ ranging from
$0.2$ to $0.9$. The qualitative consequences are hardly amended by a
change of $a$, though the absolute amount of the diphoton yield
depends on $a$.

We should remark that the choice of CASE-II may have a relation to the
Brown-Rho scaling hypothesis in dense matter~\cite{bro91},
\textit{i.e.}~$m_\sigma^\ast/m_\sigma\simeq f_\pi^\ast/f_\pi\simeq
\Phi(\rho_{\mathrm{E}})$ where $\rho_{\mathrm{E}}$ is the baryon
number density corresponding to the chemical potential
$\mu_{\mathrm{E}}$. CASE-II implies the choice of
$\Phi(\rho_{\mathrm{E}})=1/2$. Taking into account the fact that the
partial restoration of the chiral symmetry might be experimentally
observed in nuclei~\cite{bon96,hat99}, we can anticipate that such
reduction up to the half may happen in the vicinity of CEP.

\section{spectral functions near the critical end point}
\label{sec:spect}

From the experimental point of view, functional forms of the meson
masses and the condensate assigned by Eqs.~(\ref{eq:mass}) and
(\ref{eq:fpi}) are not sufficient to describe physical decay
processes, that is, the decay width and the amplitude must be
evaluated. These informations are contained in the spectral functions.
Actually the spectral functions in the $\sigma$-channel are necessary
to estimate the diphoton emission rates. Therefore, in this section,
we are going to draw our knowledge on the spectral functions in the
$\sigma$-channel, limiting our consideration to the case with the
spatial momentum fixed at zero. The spectral function in the
$\sigma$-channel is defined as
\begin{equation}
 \rho_\sigma(p)=-\frac{1}{\pi}\mathrm{Im}D_\sigma^{\mathrm{R}}(p)
  =-\frac{1}{\pi}\cdot\frac{\mathrm{Im}\Pi_\sigma^{\mathrm{R}}(p)}
  {(p^2-m^2-\mathrm{Re}\Pi_\sigma^{\mathrm{R}})^2+(\mathrm{Im}
  \Pi_\sigma^{\mathrm{R}}(p))^2}.
\label{eq:spectral}
\end{equation}
$D_\sigma^{\mathrm{R}}(p)$ and $\Pi_\sigma^{\mathrm{R}}(p)$ are the
retarded Green's function and the retarded self-energy in the
$\sigma$-channel, respectively.
In principle, all we have to do is calculate the self-energy
$\Pi_\sigma^{\mathrm{R}}(p)$ near the critical point as performed,
for example, in Ref.~\cite{chi98}.

\begin{figure}
\includegraphics[width=7cm]{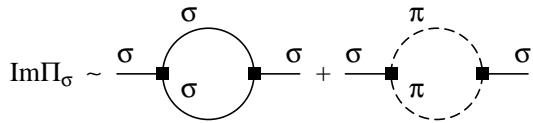}
\caption{Diagrams for the self-energy with non-vanishing
imaginary-part up to the one-loop order. Boxes stand for the
full-vertices inferred from the effective potential.}
\label{fig:diag}
\end{figure}

In place of performing some resummation procedures, we calculate only
the imaginary-part of the self-energy using the propagators with the
parametrized masses. One-loop diagrams shown in Fig.~\ref{fig:diag}
describe the decay processes $\sigma\rightarrow2\sigma$ and
$\sigma\rightarrow2\pi$. We would emphasize that this is not a simple
perturbative expansion but rather a resummed calculation since the
propagator incorporates the interaction in the mean-field of masses
given by $m_\pi(T)$ and $m_\sigma(T)$. Also the full-vertices
represented by boxes in Fig.~\ref{fig:diag} can be inferred from the
effective potential (Appendix~\ref{app:vertex} for detail arguments).

Here we limit ourselves to the case in the back-to-back kinematics,
that is, diphoton emission with zero spatial momentum. Then we obtain
\begin{align}
 \mathrm{Im}\Pi_\sigma^{\mathrm{R}}(\omega,\vec{0}) =
 & -\frac{V_{\sigma\sigma\sigma}^2}{32\pi}\sqrt{1-\frac{4m_\sigma^2}
  {\omega^2}}\Bigl(1+2n_{\mathrm{B}}(\omega/2)\Bigr)\,\theta(\omega-
  2m_\sigma) \notag\\
 & \qquad -\frac{3V_{\sigma\pi\pi}^2}{32\pi}\sqrt{1-\frac{4m_\pi^2}
  {\omega^2}}\Bigl(1+2n_{\mathrm{B}}(\omega/2)\Bigr)\,\theta(\omega-
  2m_\pi),
\end{align}
where $n_{\mathrm{B}}(\omega/2)$ stands for the usual Bose
distribution function and $\theta(\omega-2m)$ is the step function.
The expressions for the vertices $V_{\sigma\sigma\sigma}$ and
$V_{\sigma\pi\pi}$ are given in terms of $m_\pi(T)$, $m_\sigma(T)$,
and $f_\pi(T)$ in Eq.~(\ref{eq:vertices}) in
Appendix~\ref{app:vertex}.

We have neglected processes involving quark (nucleon) loops such as
$\sigma\rightarrow q\bar{q}$ ($\sigma\rightarrow N\bar{N}$) because of
physical and technical reasons. First of all, it should be negligible
at CEP since the constituent quark is still heavy ($\sim
100\;\mathrm{MeV}$), while the $\sigma$ meson becomes light.
Furthermore, the inclusion of quarks needs the wave-function
renormalization, in which we cannot avoid technical complications and
ambiguities.

It is worth noting that the parametrized mass is the curvature of the
effective potential, \textit{i.e.}~the screening mass. The pole mass
appearing in the calculation of the self-energy might be different
from the parametrized one due to the Lorentz anisotropy at finite
temperature and/or density. In Ref.~\cite{boy02} it is clarified that
the anisotropy results in the dynamical critical exponent
$z\simeq1+1/27$ for the $\phi^4$-theory slightly above
$T_{\mathrm{c}}$ at zero density. The underlying physical contents of
massless $\sigma$ mesons near CEP are effectively the same as the
critical phenomena described by the $\phi^4$-theory. Thus we
extrapolate the results of Ref.~\cite{boy02} and suppose that the
effect of the dynamical critical exponent is small enough to be
negligible even below $T_{\mathrm{c}}$ at finite density. This
presumption of neglecting the dynamical critical exponent formally
corresponds to a specific choice of renormalization conditions like
the fastest apparent convergence (FAC) condition adopted in
Hatree-type resummations~\cite{chi98}, in which the coupling constant
contained in models should depend on the temperature, in principle.

Also we briefly comment upon the results of Ref.~\cite{sca01}. The
authors argued that the pole mass of the $\sigma$ meson within the
NJL-model remains finite even right at CEP in the leading order of the
$1/N_{\mathrm{c}}$ expansion. This is not because of the treatment of
the zero-point energy as discussed by the authors, but because of the
lack of infrared dynamics. We have found that the massless $\sigma$
loops dependent on the external momentum are essential ingredients to
describe the critical properties~\cite{fuk01}.

Now that we have the imaginary-part of the self-energy, we can
construct the real-part from its imaginary-part via the dispersion
relation, \textit{i.e.}
\begin{equation}
 \mathrm{Re}\Pi_\sigma^{\mathrm{R}}(\omega,\vec{0}) =\frac{1}{\pi}
  \mathcal{P}\int_{s_0}^\infty\mathrm{d}s\,\frac{\mathrm{Im}
  \Pi_\sigma^{\mathrm{R}}(\sqrt{s},\vec{0})}{s-\omega^2}
  +\text{(subtractions)},
\label{eq:dispersion}
\end{equation}
apart from subtraction factors which are to be determined by the
renormalization conditions, in principle. $\mathcal{P}$ stands for the
prescription of Cauchy's principal value. As for the subtractions,
meson one-loops need only one subtraction, which is absorbed in the
mass-renormalization. We can fix the subtraction factor without
ambiguity by demanding the $\sigma$ mass to be given by
Eq.~(\ref{eq:mass}), neglecting the effect of the dynamical critical
exponent. If quark loops were taken into account, they would need two
subtractions in the one-loop order, which are absorbed not only in the
mass renormalization but in the wave-function renormalization also.
This makes the actual computation quite intricate, though the
quantitative behaviors are essentially described by meson loops only.

\begin{figure}
\includegraphics[width=7cm]{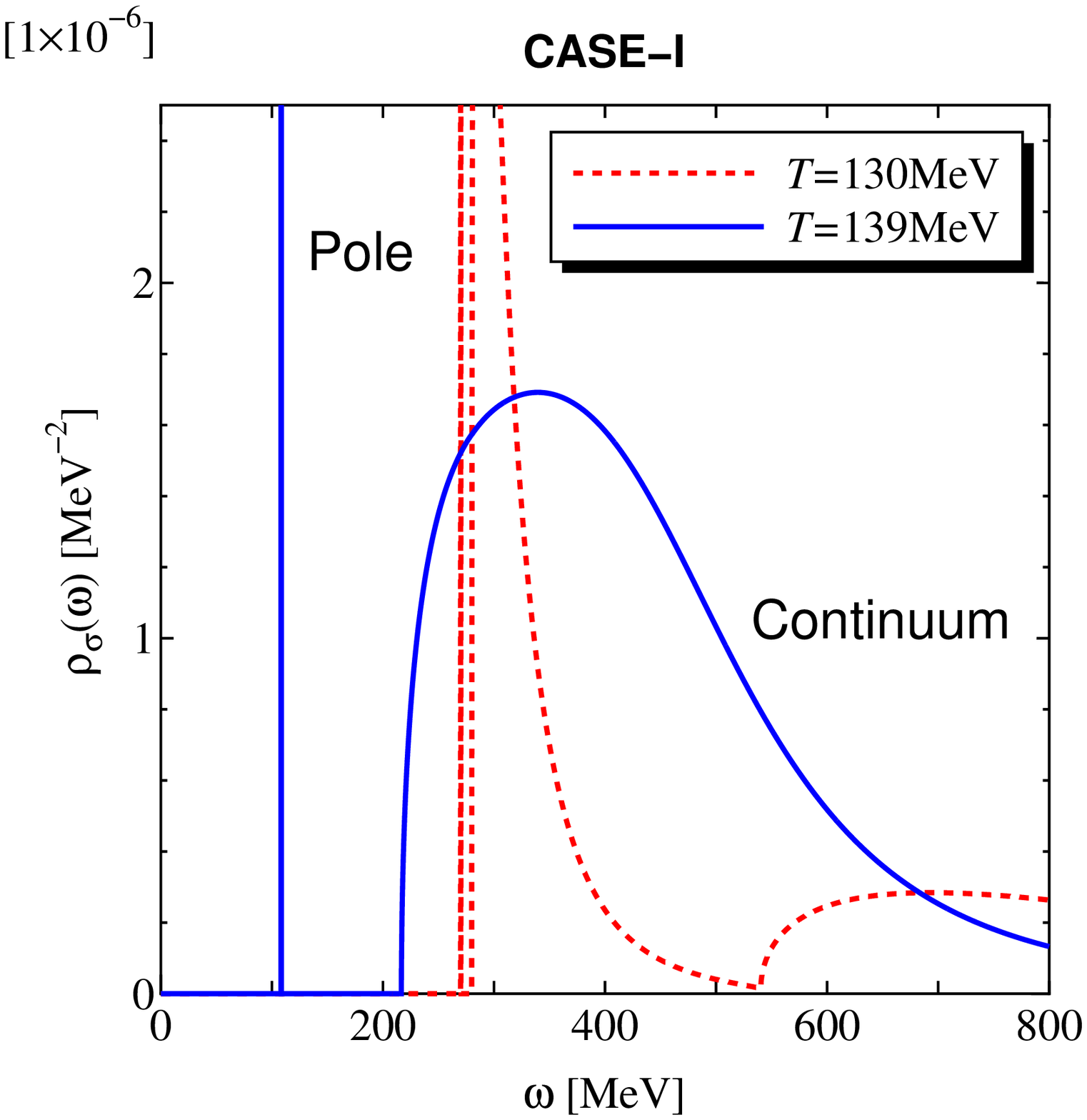}\hspace{5mm}
\includegraphics[width=7cm]{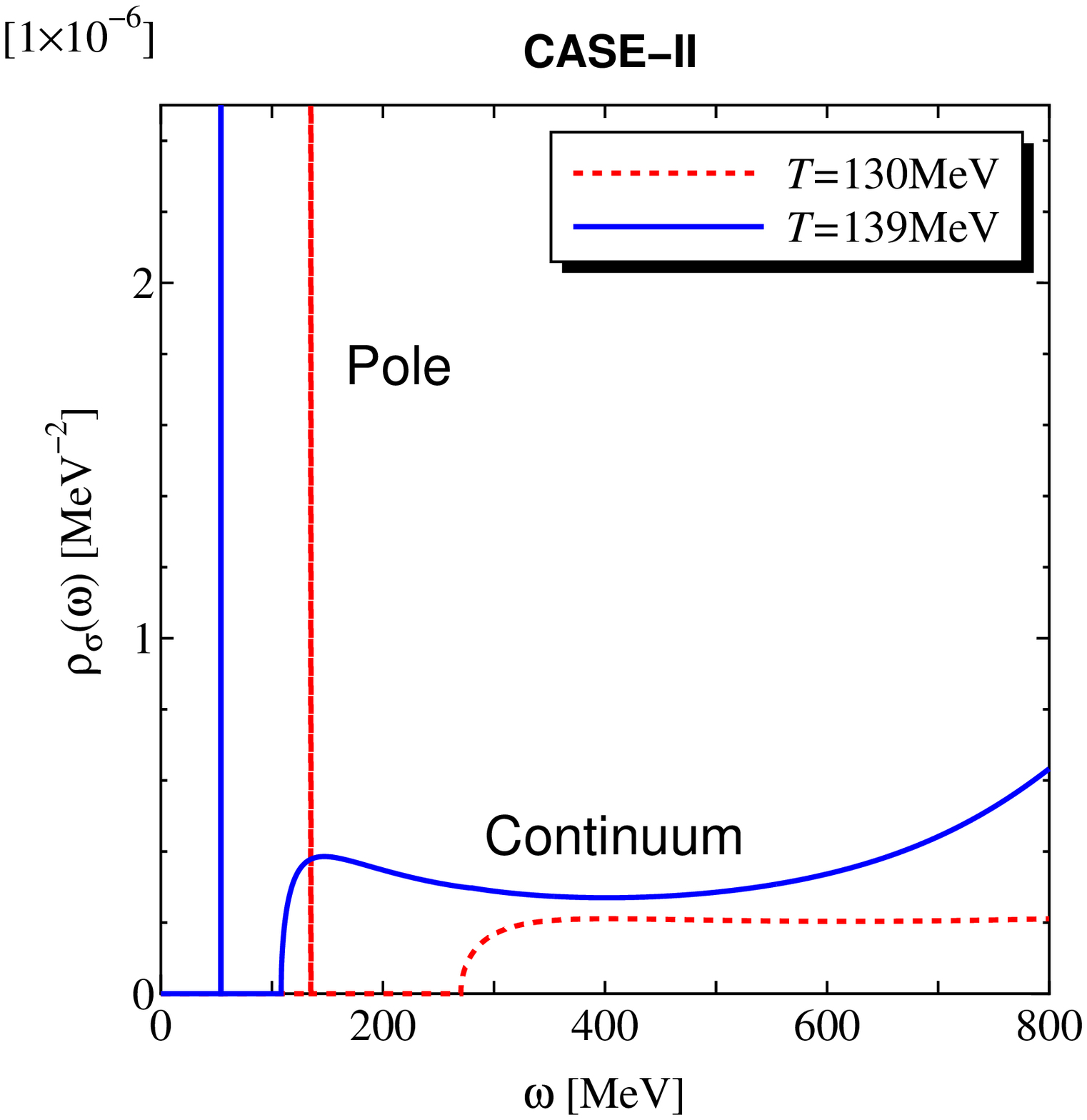}
\caption{Spectral functions in the $\sigma$-channel near CEP located
at $(\mu_{\mathrm{E}}=700\:\mathrm{MeV},
T_{\mathrm{E}}=140\:\mathrm{MeV})$. The left figure is for the
parameter set of CASE-I ($\Phi(\rho_{\mathrm{E}})=1$) and the right
for that of CASE-II ($\Phi(\rho_{\mathrm{E}})=1/2$).}
\label{fig:spectral}
\end{figure}

Then we can evaluate the spectral functions in the $\sigma$-channel by
substituting $\mathrm{Im}\Pi_\sigma^{\mathrm{R}}(\omega)$ and
$\mathrm{Re}\Pi_\sigma^{\mathrm{R}}(\omega)$ for the
definition of Eq.~(\ref{eq:spectral}). $m^2$ in
Eq.~(\ref{eq:spectral}) is irrelevant since it can be absorbed in the
subtraction factor. Throughout this analysis we set the location of
CEP at $(\mu_{\mathrm{E}}=700\:\mathrm{MeV},\:
T_{\mathrm{E}}=140\:\mathrm{MeV})$ as suggested by
Refs.~\cite{ber00,fod02}. The resulting spectral functions are shown
in Fig.~\ref{fig:spectral} for the cases of $T=130\:\mathrm{MeV}$ and
$T=139\:\mathrm{MeV}$. It is clear from the shown spectral
functions that the dominant contributions in the low energy region
arise from the light $\sigma$ pole and the continuum for the decay
process $\sigma\rightarrow2\sigma$. The threshold of the continuum
contribution lies in the point $\omega=2m_\sigma(T)$. We note that the
process $\sigma\rightarrow2\pi$ has only slight effects on the spectral
function at $T=139\:\mathrm{MeV}$ simply because the full-vertex
$V_{\sigma\pi\pi}$ becomes decreasing when the $\sigma$ mass
approaches the pion mass, as can be seen in Eq.~(\ref{eq:vertices}).

\begin{figure}
\includegraphics[width=7cm]{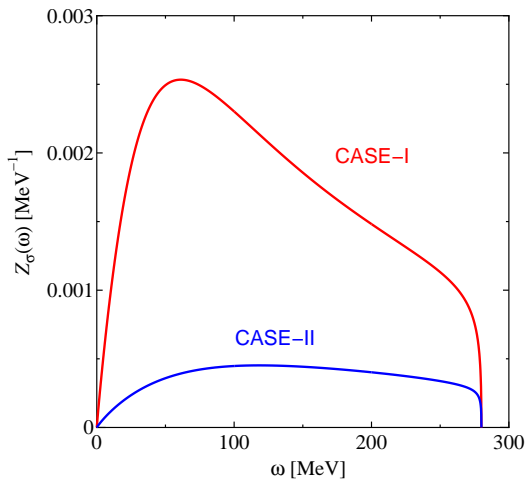}
\caption{Strength of the pole contributions.}
\label{fig:z}
\end{figure}

To investigate the spectral enhancement in a more quantitative sense,
we define the strength of the pole contribution as
\begin{equation}
 Z_\sigma(\omega)=\frac{1}{2\omega}\biggl(1-\frac{\mathrm{d}}
  {\mathrm{d}\omega^2}\Pi_\sigma^{\mathrm{R}}(\omega)\biggr)^{-1}
  \biggr|_{T=T_\omega},
\label{eq:z}
\end{equation}
where $T_\omega$ is the temperature at which
$m_\sigma(T=T_\omega)=\omega$ is satisfied. Then the pole part of the
spectral function can be written as
$\rho_\sigma^{\mathrm{(sing)}}(\omega)=Z_\sigma(\omega)
\delta(\omega-m_\sigma(T))$. In Fig.~\ref{fig:z} we plot the strength
of the pole contribution as a function of $\omega$. As is clear from
the figure, the strength becomes vanishing at $\omega=0$. This
property is understood immediately from the dispersion integral of
Eq.~(\ref{eq:dispersion}). The differentiation of the integral
diverges infraredly ($s\sim0$) as $\omega\rightarrow0$. As a result of
the infraredly diverging denominator, the spectral function becomes
zero at $\omega=0$, which is consistent with the general property of
the spectral function $\rho(\omega)=-\rho(\omega)$, for
mesons.\footnote{The author thanks T.~Hatsuda for pointing it out.}
We will discuss this behavior of the pole strength in connection with
the diphoton observation later in Sec.~\ref{sec:diphoton}.

\section{the diphoton emission rate}
\label{sec:diphoton}

One of the most promising candidates to see the spectral changes in
the $\sigma$-channel is the invariant-mass spectrum of diphoton
emission from the $\sigma$ meson. The emission rate per unit
space-time volume is given by (Appendix~\ref{app:emit} for the
derivation)~\cite{chi98b}
\begin{equation}
 \frac{\mathrm{d}N}{\mathrm{d}^4x\mathrm{d}^4p}
  \biggr|_{p=(\omega,\vec{0})}=\frac{|\omega^2
  g_{\sigma\gamma\gamma}(\omega)|^2}{(2\pi)^4}\cdot\frac{\rho_\sigma
  (\omega,\vec{0})}{\mathrm{e}^{\beta\omega\cosh\theta}-1},
\label{eq:multiplicity}
\end{equation}
with the effective coupling $g_{\sigma\gamma\gamma}$ of the decay
process $\sigma\rightarrow2\gamma$. Although the expression is
proportional to $\omega^4$, the singularities in
$g_{\sigma\gamma\gamma}(\omega)$ just compensate for $\omega^4$ in the
presence of a thermal medium and $\omega^2
g_{\sigma\gamma\gamma}(\omega)$ takes a finite value in the limit of
$\omega\rightarrow0$. The actual evaluation of
$g_{\sigma\gamma\gamma}(\omega)$ as a function of $\omega$ is
illustrated in Appendix~\ref{app:coupling}. $\theta$ is the fluid
rapidity related to the fluid velocity $v$ by $\theta=\mathrm{arctanh}
v$ arising from the Lorentz boost.

\begin{figure}
\includegraphics[width=7cm]{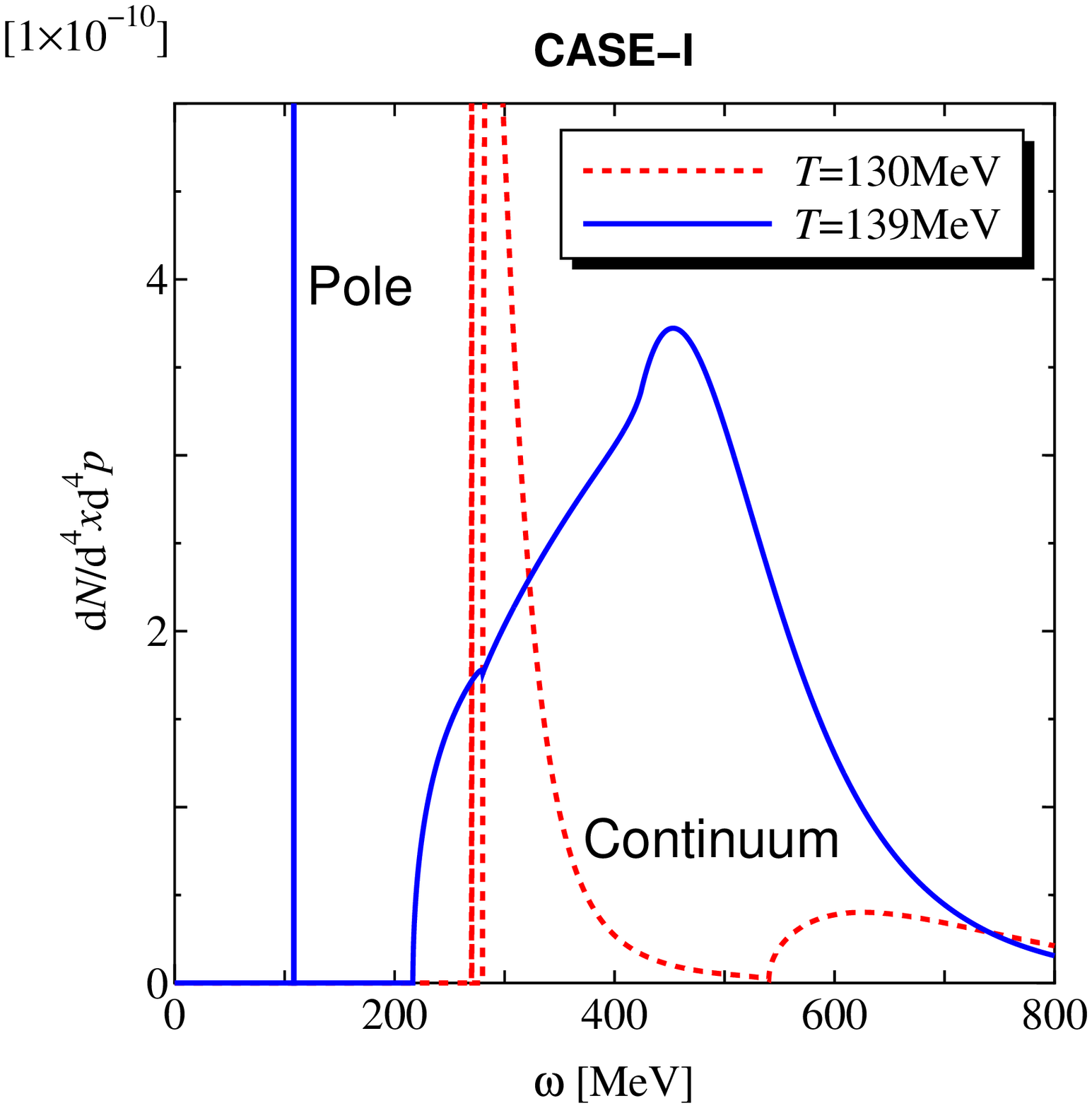}\hspace{5mm}
\includegraphics[width=7cm]{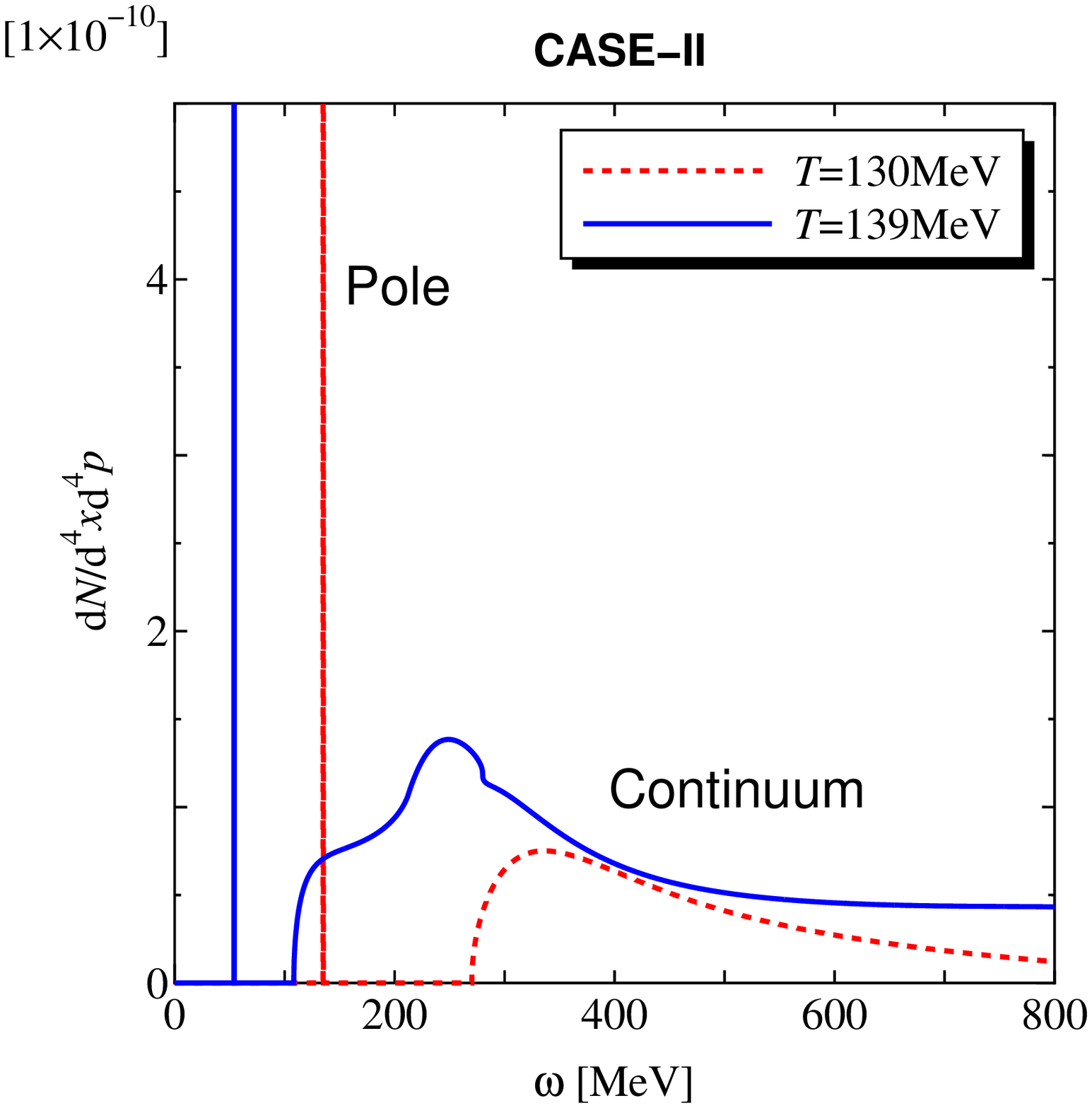}
\caption{Diphoton emission rates for $\theta=0$. The left figure is
for the parameter set of CASE-I ($\Phi(\rho_{\mathrm{E}})=1$) and the
right for that of CASE-II ($\Phi(\rho_{\mathrm{E}})=1/2$).}
\label{fig:diphoton}
\end{figure}

Since they are proportional to each others, the gross features of the
multiplicity are the same as those of the spectral function. The
dominant contribution, in fact, comes from the $\sigma$ pole when the
system lies in a state close to CEP. The results are presented in
Fig.~\ref{fig:diphoton} for $\theta=0$, in accord with the spectral
functions in Fig.~\ref{fig:spectral}, respectively.

We must take account of the space-time evolution of the system for the
calculations to be compared with experimental outputs. Here we adopt
the simplest situation for the space-time history, though it is not
quantitatively realistic but qualitatively acceptable.

First of all, we fix the baryon chemical potential at
$\mu=\mu_{\mathrm{E}}$ during the evolution until freeze-out. This
approximation makes the entropy per baryon deviate at most $\sim10\%$
depending on the nucleon mass around CEP when the equation of state
for the ideal gas is used. Considering that neither the nucleon mass
nor the equation of state is precisely known around CEP, this
approximate path with $\mu=\mu_{\mathrm{E}}$ is the best we can do for
the qualitative study.

Secondly, following the analysis presented in Ref.~\cite{cle87}, we
presume that the hydrodynamic evolution would obey the Bjorken's
scaling solution~\cite{bjo83}. In a quantitative sense the transverse
expansion should be taken into account and moreover the boost
invariance might be unsatisfied under the experimental conditions
where CEP is concerned. Nevertheless, the analysis based on the
scaling solution will provide us with the qualitative pattern of the
diphoton spectrum when CEP is passed through.

Then the invariant-mass spectrum with the transverse mass
$M_{\mathrm{t}}$, the transverse momentum $p_{\mathrm{t}}$ and the
momentum rapidity $Y$ of diphotons fixed is given by
\begin{equation}
 \frac{1}{\text{(Area)}}\cdot\frac{\mathrm{d}N}{\mathrm{d}
  M_{\mathrm{t}}^2\mathrm{d}^2 p_{\mathrm{t}}\mathrm{d}Y}
  \biggr|_{p_\mathrm{t}=Y=0} =\frac{1}{2}
  \int_{\tau_\mathrm{i}}^{\tau_\mathrm{f}}\mathrm{d}\tau\,\tau
  \int\mathrm{d}y\,
  \frac{\mathrm{d}N}{\mathrm{d}^4x\mathrm{d}^4p}
  \biggr|_{p=(M,\vec{0})},
\end{equation}
where (Area) is the constant transverse area of the hadronic gas and
$y$ is the space-time rapidity which is equal to the fluid rapidity
$\theta$ in Bjorken's scaling solution. $\tau_{\mathrm{i}}$ and
$\tau_{\mathrm{f}}$ are the initial and the final time of the
evolution; $\tau_{\mathrm{i}}$ is taken as $1\:\mathrm{fm}$ as usual
and $\tau_{\mathrm{f}}$ is defined as the time when the system is
cooled down below the freeze-out temperature. We fix the freeze-out
temperature around CEP as $T_{\mathrm{f}}=120\:\mathrm{MeV}$ according
to Ref.~\cite{ber00}. In order to accomplish the integration with
respect to $\tau$, we need the temperature at given $\tau$. That is
determined by the following argument, as discussed in
Ref.~\cite{reh97}. Bjorken's scaling solution leads to the entropy
density as a function of time as
\begin{equation}
 \frac{s(\tau)}{s(\tau_{\mathrm{i}})}=\frac{\tau_{\mathrm{i}}}{\tau},
\end{equation}
from which we can read the temperature once we have a relation
connecting the entropy density to the system temperature. For the
ideal gases, the entropy density is given by
\begin{equation}
 s_{\mathrm{H}}(T)=3\cdot\frac{4\pi^2}{90}T^3+4\cdot\frac{\partial}
  {\partial T}\biggl\{T\int\frac{\mathrm{d}^3k}{(2\pi)^3}\Bigl(
  \ln\bigl[1+\mathrm{e}^{-\beta(E_k-\mu_{\mathrm{E}})}\bigr]+\ln
  \bigl[1+\mathrm{e}^{-\beta(E_k+\mu_{\mathrm{E}})}\bigr]\Bigr)
  \biggr\},
\end{equation}
for the hadronic gas composed of pions $\pi^\pm$, $\pi^0$, and
nucleons $p$, $n$. Pions can be treated as massless particles, while
$E_k=\sqrt{k^2+M_{\mathrm{n}}^{\ast2}}$ is the energy of nucleons with
the mass $M_{\mathrm{n}}^\ast$ in a medium. For the quark-gluon plasma
composed of massless gluons and quarks with two flavors $u$, $d$ at
quark chemical potential $\mu_{\mathrm{E}}/3$, the entropy density
is given by
\begin{equation}
 s_{\mathrm{Q}}(T)=\frac{74\pi^2}{45}T^3+2T\biggl(
  \frac{\mu_{\mathrm{E}}}{3}\biggr)^2.
\end{equation}
In the intermediate temperature we can smoothly interpolate between
them as follows:
\begin{equation}
 s(T)=\frac{1}{2}\bigl(s_{\mathrm{Q}}(T)+s_{\mathrm{H}}(T)\bigr)
  +\frac{1}{2}\bigl(s_{\mathrm{Q}}(T)-s_{\mathrm{H}}(T)\bigr)\tanh
  \biggl(\frac{T-T_{\mathrm{E}}}{\Delta T}\biggr),
\label{eq:entropy}
\end{equation}
where $\Delta T$ controls the strength of the (deconfinement)
transition. It is worth noting that the change of the entropy density
is mainly attributed to the liberation of the color degrees of freedom
and thus it has little to do with the chiral phase transition, at
least in principle. There remain subtleties in this respect because no
well-defined indicator of confinement is established so
far~\cite{fuk02}. If the deconfinement transition is really related to
the chiral dynamics, as often said, the change of the entropy density
at CEP (\textit{i.e.}~the terminal point of the \textit{first-ordered}
transition) could be almost discontinuous, that is, $\Delta T$ is
small or zero in effect. Here we set $\Delta T=1\:\mathrm{MeV}$.

\begin{figure}
\includegraphics[width=7cm]{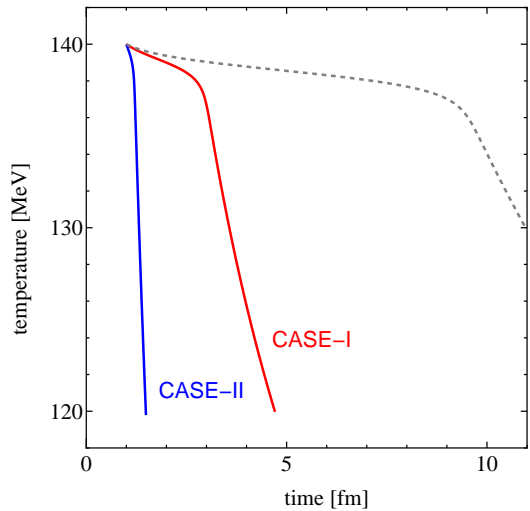}
\caption{The temperature as a function of time derived from the ansatz
given by Eq.~(\ref{eq:entropy}) with the initial condition
$T_{\mathrm{i}}=T_{\mathrm{E}}=140\:\mathrm{MeV}$ at
$\tau_{\mathrm{i}}=1\:\mathrm{fm}$.}
\label{fig:temperature}
\end{figure}

The temperature as a function of time is shown in
Fig.~\ref{fig:temperature} with the initial condition
$T_{\mathrm{i}}=T_{\mathrm{E}}=140\:\mathrm{MeV}$ at
$\tau_{\mathrm{i}}=1\:\mathrm{fm}$. We take this initial condition, in
which the gas is equilibrated right at CEP, for we have parametrized
the masses below $T_{\mathrm{E}}$ and have no idea about the masses
above CEP. The dotted curve stands for the temperature evolution in
the case when the nucleons are infinitely heavy. In CASE-I and CASE-II
the nucleon mass $M_{\mathrm{n}}^\ast$ is chosen in accord with the
Brown-Rho scaling hypothesis, that is,
$M_{\mathrm{n}}^\ast\simeq1\:\mathrm{GeV}$ for CASE-I and
$M_{\mathrm{n}}^\ast\simeq0.5\:\mathrm{GeV}$ for CASE-II.

We note that the ansatz given by Eq.~(\ref{eq:entropy}) is implicitly
based on a specific choice of the initial entropy density
$s(T_{\mathrm{E}})=(s_{\mathrm{H}}(T_{\mathrm{E}})+s_{\mathrm{Q}}
(T_{\mathrm{E}}))/2$. The gentle slope of temperature until
$\tau\sim3\:\mathrm{fm}$ ($1.2\:\mathrm{fm}$) in CASE-I (CASE-II) is
caused by the difference of the entropy density
$s(T_{\mathrm{E}})-s_{\mathrm{H}}(T_{\mathrm{E}})
=(s_{\mathrm{Q}}(T_{\mathrm{E}})-s_{\mathrm{H}}(T_{\mathrm{E}}))/2$.
Although we will focus our arguments only on the case of the
ansatz~(\ref{eq:entropy}), other forms of interpolation will be
necessary when we look into the case with different initial conditions
for the entropy density.

\begin{figure}
\includegraphics[width=7cm]{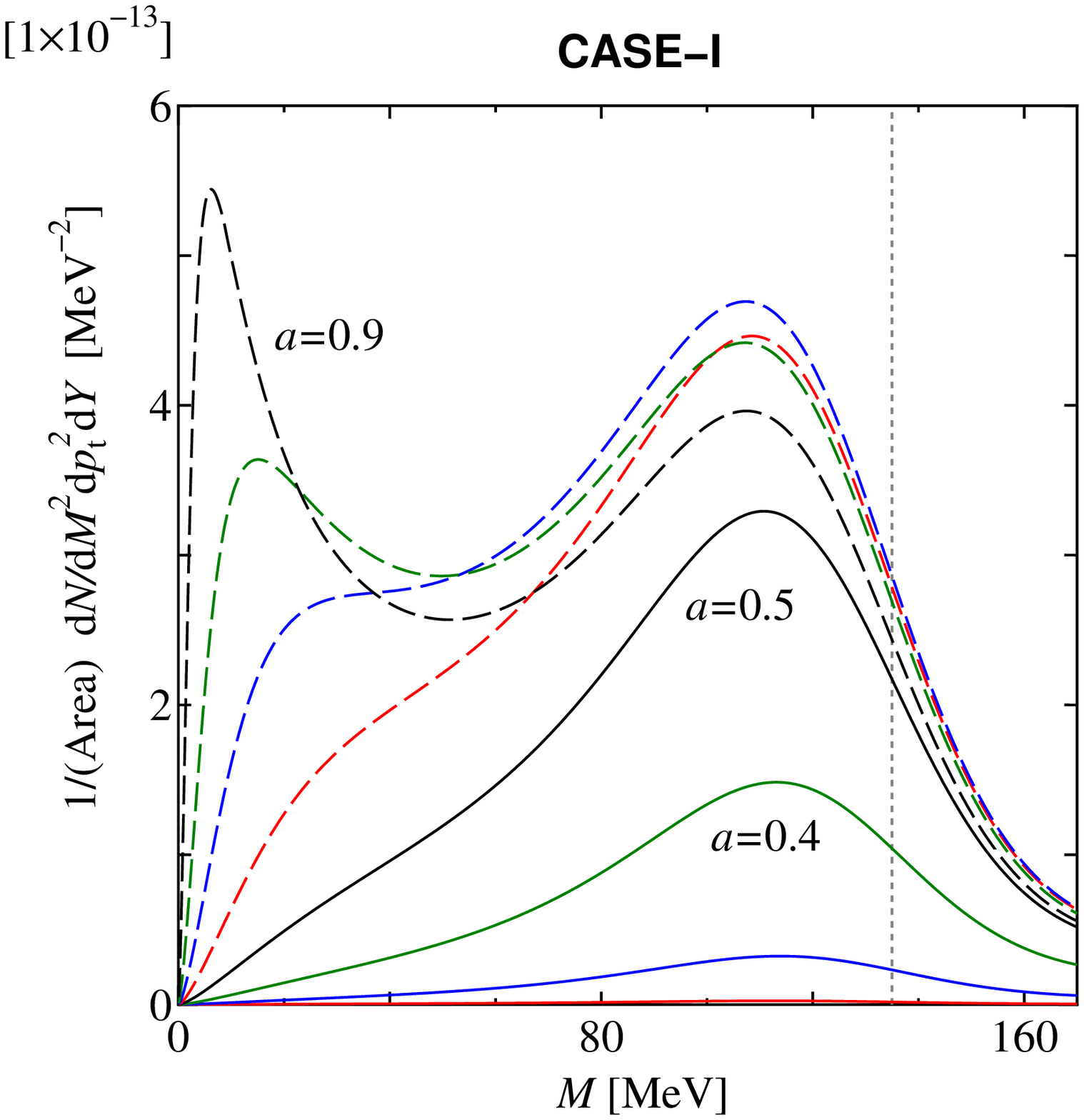}\hspace{5mm}
\includegraphics[width=7cm]{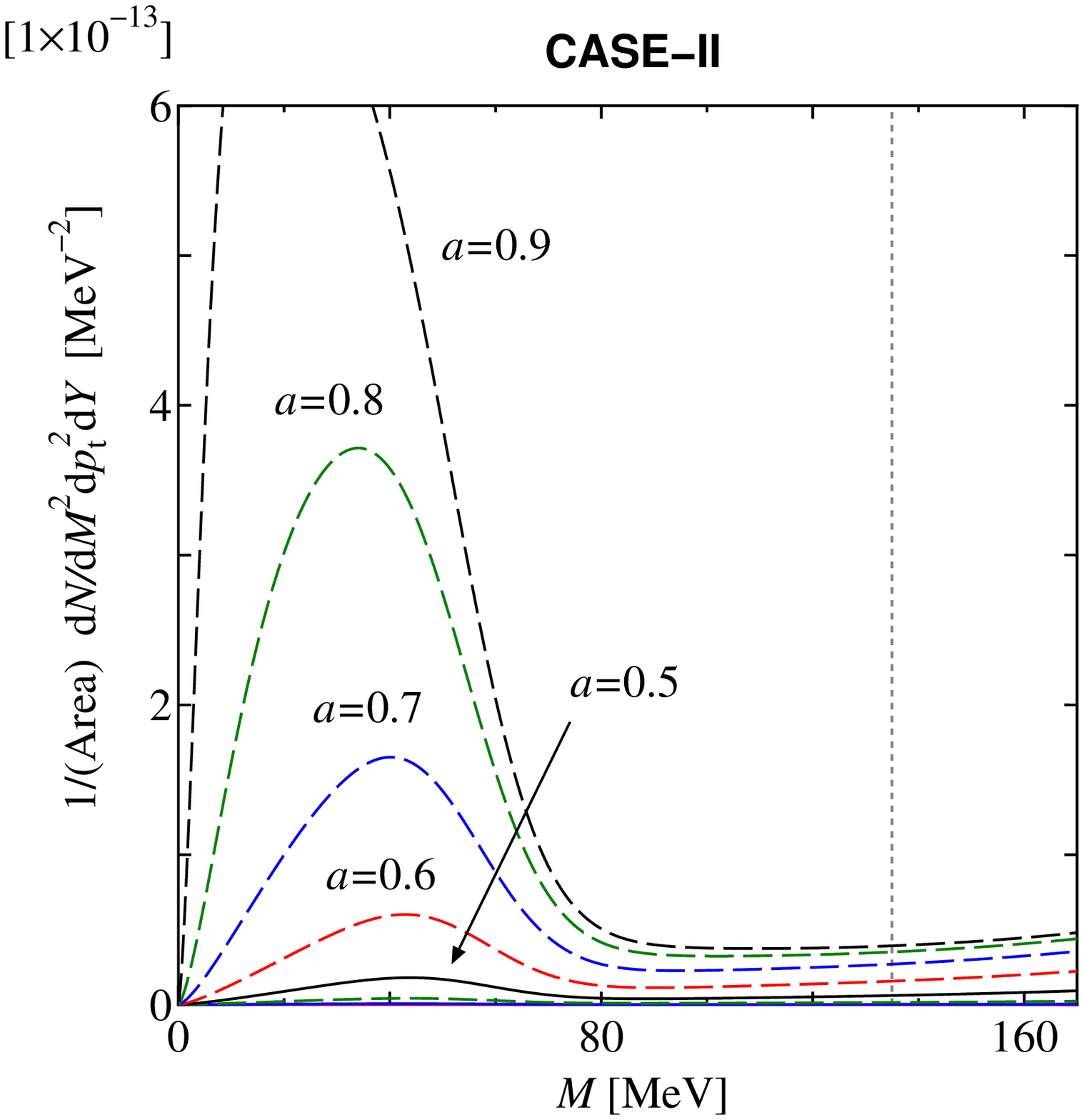}
\caption{The multiplicity of diphoton emission near CEP. The left
figure is for the parameter set of CASE-I
($\Phi(\rho_{\mathrm{E}})=1$) and the right for that of CASE-II
($\Phi(\rho_{\mathrm{E}})=1/2$). The dotted line indicates the
position of the pion mass $m_{\pi^0}=135\:\mathrm{MeV}$.}
\label{fig:final}
\end{figure}

Fig.~\ref{fig:final} is our final result for the diphoton emission
when the system passes through CEP. Because the continuum contribution
turns out to be $2\sim3$ orders of magnitude smaller than the pole
contribution, we have plotted only the pole contribution with a
variety of $a$ ranging from 0.2 (bottom) to 0.9 (top). We cut the
invariant-mass $M$ shown in the figures up to $170\:\mathrm{MeV}$, for
the $\sigma$ mass at the freeze-out temperature is around
$170\:\mathrm{MeV}$ in CASE-II.

At a glance we notice a characteristic property that the multiplicity
approaches zero as $M\rightarrow0$ and has a peak around
$M=50\sim120\:\mathrm{MeV}$. Although we have expected that the
presence of massless $\sigma$ mesons would induce more and more
significant enhancement in the multiplicity as $M$ gets smaller and
smaller, the resulting multiplicity in the low energy region is
reduced altogether. This suppression is because of the amplitude, or
the strength of the pole contribution, which is given by
Eq.~(\ref{eq:z}).

The location of the peak actually depends on $m_\sigma^\ast$,
$f_\pi^\ast$ (the $\sigma$ mass and the pion decay constant at
$(\mu=\mu_{\mathrm{E}},\: T=0)$), and the evolution of temperature as
a function of time. In any case the multiplicity starts from zero at
$M=0$. As the energy gets larger, the amplitude rapidly increases and
at the same time the space-time volume in which the thermal system
stays is reduced by the steeper slope of temperature. Consequently the
peak observed in Fig.~\ref{fig:final} is produced in somewhat general
situations due to the behavior of the amplitude and the time
evolution. Thus we can anticipate that the peak found in our analysis
appears in different physical conditions (different choices of
$m_\sigma^\ast$, $f_\pi^\ast$, and the hydrodynamic evolution) as long
as the system passes through CEP, though its location is altered
according to those conditions.

We would remark that the annihilation process
$\pi^+\pi^-\rightarrow2\gamma$ makes no background in such a low
energy region below the threshold of two pions. In the scenario of the
spectral enhancement in the chiral restoration at zero chemical
potential, the annihilation process brings about a huge background
comparable to the desired signal, which crucially smears the spectral
changes~\cite{chi98}. In contrast, the main background for the CEP
peak comes from the electro-magnetic decay into two photons from
$\pi^0$ in a hot medium and also from $\pi^0$ emitted after
freeze-out. In the present analysis the pion mass is fixed at a
constant value independently of temperature. In addition to the pole
contribution at $M=m_{\pi^0}=135\:\mathrm{MeV}$, the spectral
functions in the pion-channel have a continuum contribution in the low
energy region below the pion mass because of the mass difference
between pions and $\sigma$ mesons. However, this pion continuum gives
rise to so small contribution, which is at most comparable to the
continuum contribution in the $\sigma$-channel, that it is
sufficiently negligible.

Although the invariant-mass spectrum has a characteristic shape with a
peak whose location is separable from the pion pole, the combinatorial
background of photons from $\pi^0\rightarrow2\gamma$ will be crucial
for experimental observations. The invariant-mass distribution of
photon pairs from separate sources is the origin of the combinatorial
background. In principle, however, the combinatorial background can be
subtracted by means of the event-mixing method. Thus the accuracy
necessary to identify the peak originating from $\sigma$ mesons is
roughly estimated by the comparison between the number of photons from
$\sigma$ mesons and the number of photons from $\pi^0$ which might
pair with photons from $\sigma$ mesons in the construction of the
invariant-mass.

Let us roughly estimate the accuracy necessary to identify the
diphoton peak from the $\sigma$ meson in the present formulation,
though the analysis should not be reliable beyond a qualitative sense
and the quantitative results may be significantly changed under
different conditions for the choice of parameters, the initial values
of the temperature and the chemical potential, the equation of state,
the hydrodynamic evolution, and so on.

The $\pi^0$ yield in the present condition can be estimated by the
integration of the Bose distribution function on the freeze-out
surface as~\cite{cse98}
\begin{equation}
 \frac{1}{\text{(Area)}}\frac{\mathrm{d}N_{\pi^0}}{\mathrm{d}Y}
  \simeq\frac{\zeta(3)\tau_{\mathrm{f}}T_{\mathrm{f}}^3}{\pi^2}
  \simeq 0.13\:\mathrm{fm}^{-2},
\label{eq:pions}
\end{equation}
with $\tau_{\mathrm{f}}=4.7\:\mathrm{fm}$ and
$T_{\mathrm{f}}=120\:\mathrm{MeV}$ (CASE-I) substituted. This value of
the $\pi^0$ yield seems small as compared with, for example, the
particle production observed in AGS~\cite{ahl98}. It is because we set
the initial point where the system reaches equilibrium right at CEP,
not above it. As a result the freeze-out time $\tau_{\mathrm{f}}$
becomes smaller.

Since almost all the $\pi^0$ decays via an electro-magnetic process
into two photons, the number of photons from $\pi^0$ is estimated as
twice of Eq.~(\ref{eq:pions}). On the other hand, the invariant-mass
spectrum of diphoton emission with the range up to
$p_{\mathrm{t}}\sim M\sim100\:\mathrm{MeV}$ integrated out gives
\begin{equation}
 \frac{1}{\text{(Area)}}\frac{\mathrm{d}N_{\sigma\gamma\gamma}}
  {\mathrm{d}Y}\sim 10^{-8}\:\mathrm{fm}^{-2}.
\label{eq:output}
\end{equation}
As a rough estimate, the accuracy of order $\sim10^{-7}$ to
the $\pi^0$ peak after the subtraction of the combinatorial background
is necessary to detect the clear signal from nearly massless $\sigma$
mesons. The infrared suppression of the amplitude we find in the
present paper is responsible for the smallness of the resulting
output~(\ref{eq:output}). As we stated above, however, the
quantitative results here may be considerably changed. For reliable
quantitative analyses, we must clarify the hydrodynamic properties
around CEP, which are just beginning to be investigated~\cite{asa02}.

\section{summary}
\label{sec:summary}

We investigate the spectral functions in the $\sigma$-channel near
CEP where the first-ordered transition of the chiral restoration would
terminate. The $\sigma$ pole and the continuum from the decay process
$\sigma\rightarrow2\sigma$ dominate over the spectral functions near
CEP, as expected. Our method has almost no model artifact though it
contains some parameters put by hand. In the present analysis we try
two parameter sets in the extreme cases; one is the case with no
medium effect and another is the case with a considerably large medium
effect.

Using the resulting spectral functions we have evaluated the
multiplicity of diphoton emission. Within the simplest space-time
evolution described by Bjorken's scaling solution, we present the
theoretical prediction for the diphoton emission when the system
passes through CEP in a hydrodynamic evolution. Our results show that
the diphoton multiplicity with vanishing transverse momentum has the
characteristic shape with a peak around the invariant mass
$50\sim120\:\mathrm{MeV}$. We find, contrary to expectation, that the
infrared dynamics makes the amplitude of nearly massless $\sigma$
mesons so suppressed that the CEP signal is weakened. The severe
background comes from a large peak produced by $\pi^0$ after
freeze-out.

To proceed further and employ more realistic hydrodynamics, it is
essential for the present framework to settle the parametrization in
the whole $(\mu,T)$-plane and extend it above the critical point. As
for the chiral critical point in three flavor QCD at finite
temperature~\cite{gav94}, lattice simulations prove to be powerful
instruments to search for the $t$-like and the $h$-like
directions~\cite{kar01}. Unfortunately, however, CEP located at high
density is still hard to access by means of lattice simulations.

\begin{acknowledgments}
The author, who is supported by Research Fellowships of the Japan
Society for the Promotion of Science for Young Scientists, would like
to thank H.~Fujii, T.~Hirano, T.~Matsui, and K.~Ohnishi for
discussions. He also would like to express gratitude to T.~Hatsuda for
careful reading of the manuscript and giving valuable comments.
\end{acknowledgments}

\appendix

\section{interactions near the critical end point}
\label{app:vertex}

In the presence of finite chemical potential for the baryon density,
the effective potential in terms of $\sigma$ and $\pi$ can be expanded
as
\begin{equation}
 V(\sigma,\vec{\pi})=A(\sigma^2+\vec{\pi}^2)^3-B(\sigma^2
  +\vec{\pi}^2)^2+C(\sigma^2+\vec{\pi}^2)-D\sigma,
\end{equation}
where the last term embodies explicit breaking of the chiral symmetry
due to finite quark masses. The effects arising from finite chemical
potential induce the first term of the sixth power (finite $A$) and
reduce the second term of the fourth power (small $B$). As a result
the phase transition could be first-ordered. In the vicinity of CEP,
coefficients $A$, $B$, $C$, and $D$ are determined by the following
conditions: $\partial^2 V/\partial\pi^2=m_\pi^2$ and $\partial^2
V/\partial\sigma^2=m_\sigma^2$ for the meson masses given by
Eq.~(\ref{eq:mass}), $\partial V/\partial\sigma=0$ for the
condensation given by Eq.~(\ref{eq:fpi}), and
$\partial^3V/\partial\sigma^3=0$ for CEP. We need one more condition
to determine all the coefficients away from CEP because the last
condition is peculiar just to CEP. Then we assume that $A$ can be
regarded as constant in our analysis. Actually we can expect that the
value of $A$ is more sensitive to $\mu$, which is fixed as
$\mu=\mu_{\mathrm{E}}$ throughout this paper, rather than $T$
remembering that $A$ stems from the effects of finite chemical
potential. Then it follows that
\begin{equation}
 A=\frac{m_\pi^2}{16v_0^4},\quad B=\frac{3f_\pi^2m_\pi^2}{16v_0^4}
  +\frac{m_\pi^2-m_\sigma^2}{8f_\pi^2},\quad
 C=\frac{3f_\pi^4m_\pi^2}{16v_0^4}+\frac{3m_\pi^2-m_\sigma^2}{4},
  \quad D=f_\pi m_\pi^2.
\end{equation}
leading to the full-vertices of the three-point interactions as
\begin{align}
 V_{\sigma\sigma\sigma}&\equiv\frac{\partial^3V}{\partial\sigma^3}
  =\frac{3f_\pi^3m_\pi^2}{v_0^4}-\frac{3(m_\pi^2-m_\sigma^2)}{f_\pi},
  \notag\\
 V_{\sigma\pi\pi}&\equiv\frac{\partial^3V}{\partial\sigma\partial\pi^2}
  =-\frac{m_\pi^2-m_\sigma^2}{f_\pi}.
\label{eq:vertices}
\end{align}

\section{formulation of the diphoton emission rate}
\label{app:emit}

The multiplicity of two photons (diphotons) per unit space-time volume
is given in the thermal circumstance by
\begin{equation}
 \frac{\mathrm{d}N}{\mathrm{d}^4x}=\sum_{f,i,\lambda_1,\lambda_2}
  \frac{\mathrm{e}^{-\beta E_i}}{Z(\beta)}\cdot\frac{|\langle f;
  \gamma(k_1,\lambda_1)\gamma(k_2,\lambda_2)|S|i\rangle|^2}{VT}\cdot
  \frac{\mathrm{d}^3k_1}{2\omega_1(2\pi)^3}\cdot
  \frac{\mathrm{d}^3k_2}{2\omega_2(2\pi)^3},
\end{equation}
where $Z(\beta)$ is the partition function and the $S$-matrix is
denoted by $S$. Two photons have four-momenta $(\omega_1=|k_1|,
\vec{k_1})$, $(\omega_2=|k_2|,\vec{k_2})$ and polarization vectors
$\epsilon_\mu(\lambda_1)$, $\epsilon_\mu(\lambda_2)$ respectively. The
$S$-matrix depends on the actual processes which can be expressed
through the effective Lagrangian, \textit{i.e.}
\begin{equation}
 \mathcal{L}_{\phi\gamma\gamma}=g_{\phi\gamma\gamma}\phi
  F_{\mu\nu}F^{\mu\nu},
\end{equation}
with the effective coupling constant $g_{\phi\gamma\gamma}$ of order
$\alpha_{\mathrm{e}}\simeq1/137$. In our notation $\phi$ denotes
collectively any particle which can decay into two photons such as
$\pi^0$ and $\sigma$ mesons. We will focus our attention to the case
of $\phi=\sigma$ in the present paper because we are interested in the
diphoton spectrum in the low energy region. Up to the lowest order of
the perturbation with respect to $g_{\phi\gamma\gamma}$, the
$S$-matrix can be expanded as
\begin{align}
 &\langle f;\gamma(k_1,\lambda_1)\gamma(k_2,\lambda_2)|S|i\rangle
  \notag\\
 &=-4\mathrm{i}g_{\phi\gamma\gamma}\bigl\{(k_1\cdot k_2)(
  \epsilon_1^\ast\cdot\epsilon_2^\ast)-(k_1\cdot\epsilon_2^\ast)
  (k_2\cdot\epsilon_1^\ast)\bigr\}\int\mathrm{d}^4x\,
  \mathrm{e}^{\mathrm{i}(k_1+k_2)}\langle f|\hat{\phi}(x)|i\rangle.
\end{align}
Since real (on-shell) photons are emitted, we must take the summation
with respect to the polarizations only over the transverse components.
The Ward identity, however, simplifies the manipulations, that is, we
reach the correct answer simply by taking the summation over all the
polarizations. Thus we can make use of the formula, $\sum_\lambda
\epsilon_\mu(\lambda)\epsilon_\nu^\ast(\lambda)=-g_{\mu\nu}$, to
acquire
\begin{equation}
 \frac{\mathrm{d}N}{\mathrm{d}^4x}=\mathrm{i}D_\phi^{<}(k_1+k_2)\cdot
  32|g_{\phi\gamma\gamma}|^2(k_1\cdot k_2)^2\cdot
  \frac{\mathrm{d}^3k_1}{2\omega_1(2\pi)^3}\cdot\frac{\mathrm{d}^3k_2}
  {2\omega_2(2\pi)^3},
\end{equation}
after taking the summation over states. Here, the Green's function,
\begin{equation}
 \mathrm{i}D_\phi^{<}(p)=\int\mathrm{d}^4x\,\mathrm{e}^{\mathrm{i}
  p\cdot x}\langle\hat{\phi}(0)\hat{\phi}(x)\rangle_\beta,
\end{equation}
can be related to the spectral function as~\cite{bel96}
\begin{equation}
 \mathrm{i}D_\phi^{<}(p)=\frac{2\pi\rho_\phi(p)}
  {\mathrm{e}^{\beta p_0}-1},
\end{equation}
where the spectral function is defined as Eq.~(\ref{eq:spectral}). If
the thermal medium has the flow velocity specified by $u_\mu$ then
$\beta p_0$ in the above expression should be replaced by
$\beta p\cdot u$.

In the kinematics specified by $\omega_1+\omega_2=\omega$ and
$\vec{p}=\vec{0}$ (back-to-back kinematics), we can easily perform
the integration over $\vec{k}_1$ and $\vec{k}_2$. Taking into account
the symmetry factor of identical bosons (photons), we finally reach
\begin{equation}
 \frac{\mathrm{d}N}{\mathrm{d}^4x\mathrm{d}^4p}=\frac{|\omega^2
  g_{\phi\gamma\gamma}(\omega)|^2}{(2\pi)^4}\cdot
  \frac{\rho_\phi(\omega,\vec{0})}{\mathrm{e}^{\beta\omega}-1}.
\end{equation}

\section{effective coupling constant}
\label{app:coupling}

\begin{figure}
\includegraphics[width=14cm]{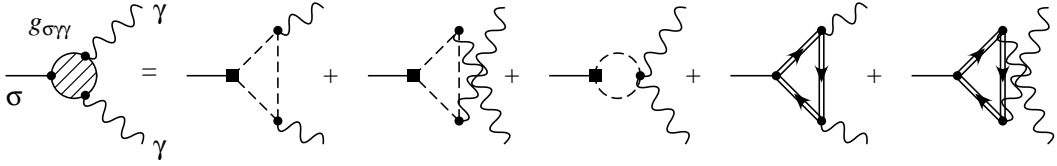}
\caption{Diagrams contributing the decay process
$\sigma\rightarrow2\gamma$ in the chiral quark model. The box
indicates the full vertex of $\sigma\rightarrow2\pi$ inferred from the
effective potential.}
\label{fig:coupling}
\end{figure}

The effective coupling constant describing the decay
$\sigma\rightarrow2\gamma$ has been calculated within the framework of
the linear sigma model at zero temperature~\cite{had88} and within the
framework of the NJL model at finite temperature~\cite{reh97}. In the
present analysis we estimate the coupling $g_{\sigma\gamma\gamma}$ by
means of the chiral quark model at finite baryon density as well as at
finite temperature. Diagrams to be calculated are shown in
Fig.~\ref{fig:coupling}. The result in the back-to-back kinematics is
written as
\begin{equation}
 g_{\sigma\gamma\gamma}(\omega)=\alpha_{\mathrm{e}}\bigl(
  t_\pi^{\text{(vac)}}(\omega)+t_\pi^{\text{(mat)}}(\omega)
  +t_q^{\text{(vac)}}(\omega)+t_q^{\text{(mat)}}(\omega)\bigr),
\end{equation}
where $\alpha_{\mathrm{e}}=e^2/4\pi$ is the electro-magnetic coupling
constant. $t_\pi^{\text{(vac)}}$ and $t_\pi^{\text{(mat)}}$ represent
the vacuum and the medium contributions of the pion loops
respectively, given by
\begin{align}
 t_\pi^{\text{(vac)}}(\omega) &=\frac{V_{\sigma\pi\pi}}{4\pi\omega^2}
  \Biggl\{1-\frac{4m_\pi^2}{\omega^2}\biggl(\sin^{-1}\frac{\omega}
  {2m_\pi}\biggr)^2\Biggr\},\notag\\
 t_\pi^{\text{(mat)}}(\omega) &=\frac{V_{\sigma\pi\pi}m_\pi^2}
  {\pi\omega^2}\Biggl\{\int_0^\infty\mathrm{d}q\frac{2q\,
  n_{\mathrm{B}}(E_q)}{E_q^2(\omega^2-4E_q^2)}\ln\biggl|
  \frac{E_q+q}{E_q-q}\biggr|\notag\\
 &\qquad\qquad -\theta(\omega-2m_\pi)\,\frac{\mathrm{i}\pi}{\omega^2}
  n_{\mathrm{B}}(\omega/2)\ln\Biggl|\frac{\omega+\sqrt{\omega^2
  -4m_\pi^2}}{\omega-\sqrt{\omega^2-4m_\pi^2}}\Biggr|\Biggr\},
\end{align}
with the notation $E_q=\sqrt{q^2+m_\pi^2}$. The vacuum and the medium
contributions, $t_q^{\text{(vac)}}$ and $t_q^{\text{(mat)}}$ come from
the quark loops, which are given by
\begin{align}
 t_q^{\text{(vac)}}(\omega) &=-\frac{5g\,m_q}{3\pi\omega^2}\Biggl\{1
  +\biggl(1-\frac{4m_q^2}{\omega^2}\biggr)\biggl(\sin^{-1}
  \frac{\omega}{2m_\pi}\biggr)^2\Biggr\},\notag\\
 t_q^{\text{(mat)}}(\omega) &=-\frac{5g\,m_q(\omega^2-4m_q^2)}{3\pi
  \omega^2}\Biggl\{\int_0^\infty\mathrm{d}q\frac{q(n_{\mathrm{F}}
  (E_q+\mu_q)+n_{\mathrm{F}}(E_q-\mu_q))}{E_q^2(\omega^2-4E_q^2)}\ln
  \biggl|\frac{E_q+q}{E_q-q}\biggr| \notag\\
 &\qquad -\theta(\omega-2m_q)\,\frac{\mathrm{i}\pi}{2\omega^2}
  (n_{\mathrm{F}}(\omega/2+\mu_q)+n_{\mathrm{F}}(\omega/2-\mu_q))\ln
  \Biggl|\frac{\omega+\sqrt{\omega^2-4m_q^2}}{\omega-\sqrt{\omega^2
  -4m_q^2}}\Biggr|\Biggr\},
\end{align}
with $E_q=\sqrt{q^2+m_q^2}$ as above. $g$ is the strength of the
Yukawa-coupling between mesons and quarks. $m_q$ denotes the
constituent quark mass. We have chosen $g=3.2$ so as to reproduce the
constituent quark mass at tree-level, that is,
$m_q=gf_\pi=300\:\mathrm{MeV}$~\cite{goc91}.

It is important to note that diagrams in Fig.~\ref{fig:coupling} have
no $\sigma$ loop since the $\sigma$ mesons are electrically neutral.
Thus  it is absolutely necessary to include the contribution from
quark loops in contrast to the calculation of the self-energy in which
the $\sigma$ contribution dominates over the quark loops.

\end{document}